\documentclass[aip,jap,reprint,amsmath,amssymb,amsfonts, floatfix, showpacs,intlimits]{revtex4-1}

\usepackage{graphicx}
\usepackage{setspace}
\usepackage{bm}

\def\<{\left\langle}
\def\>{\right\rangle}
\def\ii{{\rm i}}

\begin{document}

\title
{
Terahertz frequency spectrum analysis with a nanoscale antiferromagnetic tunnel junction
}

\author{P.~Yu.~Artemchuk}
\email{doompeter@ukr.net}
\author{O.~R.~Sulymenko}
\affiliation{Faculty of Radio Physics, Electronics and Computer Systems, Taras Shevchenko National University of Kyiv, 01601 Kyiv, Ukraine}

\author{S.~Louis}
\author{J.~Li}
\affiliation{Electrical and Computer Engineering, Oakland University, Rochester, Michigan 48309, USA}

\author{R.~Khymyn}
\affiliation{Department of Physics, University of Gothenburg, 41296 Gothenburg, Sweden}
\affiliation{NanOsc AB, 16440 Kista, Sweden}

\author{E.~Bankowski}
\author{T.~Meitzler}
\affiliation{U.S. Army TARDEC, Warren, Michigan 48397, USA}

\author{V.~S.~Tyberkevych}
\author{A.~N.~Slavin}
\affiliation{Department of Physics, Oakland University, Rochester, Michigan 48309, USA}

\author{O.~V.~Prokopenko}
\email{oleksandr.prokopenko@gmail.com}
\affiliation{Faculty of Radio Physics, Electronics and Computer Systems, Taras Shevchenko National University of Kyiv, 01601 Kyiv, Ukraine}

\begin{abstract}
A method to perform spectrum analysis on low power signals between $0.1$ and 10~THz is proposed.
It
utilizes a nanoscale antiferromagnetic tunnel junction (ATJ) that produces an oscillating tunneling anisotropic magnetoresistance, whose frequency is dependent on the magnitude of an evanescent spin current.
It is first shown that the ATJ oscillation frequency can be tuned linearly with time.
Then, it is shown that the ATJ output is highly dependent on matching conditions that are highly dependent on the dimensions of the dielectric tunneling barrier.
Spectrum analysis can be performed by using an appropriately designed ATJ, whose frequency is driven to increase linearly with time, a low pass filter, and a matched filter.
This method of THz spectrum analysis, if realized in experiment, will allow miniaturized electronics to rapidly analyze low power signals with a simple algorithm.
It is also found by simulation and analytical theory that for an ATJ with a $0.09~\mu$m$^2$ footprint, spectrum analysis can be performed over a $0.25$~THz bandwidth in just 25~ns on signals that are at the Johnson-Nyquist thermal noise floor.
\end{abstract}

\maketitle

\section{Introduction}
\label{s:Intro}

Currently, there are no existing compact technologies that are capable of rapidly performing spectrum analysis on a signal with a frequency in the bandwidth between $0.1$ and 10~THz.
This bandwidth has been dubbed the ``THz gap'' because traditional silicon electronics and traditional photonics hardware do not function effectively and thus are not capable of generating, detecting, or otherwise processing these signals \cite{Sirtori2002Nat, Kleiner2007Sci, Tonouchi2007NatPhoton, Roadmap2017JPhysD}.
In contrast, antiferromagnetic (AFM) materials show intrinsic resonant characteristics within the THz gap, and have been identified as building blocks for a new class of devices that will function at THz frequencies, as shown in many recent experimental and theoretical works \cite{Kirilyuk2010RMP, Kampfrath2011NatPhoton, Ivanov2014LowTempPhys, Jungwirth2016NatNano, Baltz2018RMP, Gomonay2010PRB, Gomonay2014LTP, Cheng2016PRL, Sinova2015RMP, Park2011NatMater, Wadley2016Sci, Khymyn2017AIPAdv, artemchuk,Sulymenko2017PRAppl, Khymyn2017SciRep, Sulymenko2018IEEEML, Khymyn2018arXiv, Sulymenko2018JAP}.
Thus far, there have been proposals for miniaturized THz frequency (TF) detectors \cite{Khymyn2017AIPAdv, artemchuk}, sources \cite{Sulymenko2017PRAppl, Khymyn2017SciRep, Sulymenko2018IEEEML}, and spiking neurons for neuromorphic applications \cite{Khymyn2018arXiv, Sulymenko2018JAP}.
In this paper, we describe how AFM materials can be used to produce a compact, simple spectrum analyzer that is functional in the THz gap.

We will show that spectrum analysis can be performed with a recently described \cite{Sulymenko2018IEEEML} antiferromagnetic tunnel junction (ATJ) in combination with a recently proposed spectrum analysis algorithm \cite{Louis2018APL}.
With this algorithm, the ATJ is used to generate TF signals whose frequency is dynamically tuned through the scanning bandwidth under the action of a ramped dc bias current.
The TF signal to be analyzed is then mixed with the dynamically tuned ATJ signal, and then processed with a low pass filter and a matched filter.
The output of the matched filter will contain the spectrum of the input TF signal encoded in time, and will have a high signal to noise ratio (SNR) that is independent of the phase difference between the input signal and the ATJ signal.
The algorithm is functional when the input TF signal has a low power, given that it is above the Johnson-Nyquist noise floor.

To demonstrate the viability of performing spectrum analysis with an ATJ, two critical areas must be investigated.
First, we must investigate of the dynamic tuning of an ATJ, and ensure that it can be tuned linearly to allow application of the spectrum analysis algorithm.
Second, we must develop a circuit model to describe the electrical behavior of an ATJ when interfacing with an external signal at THz frequencies.
Once these two tasks have been performed, the performance of the ATJ can be optimized and THz spectrum analysis can be performed.
Before we investigate these two areas, however, this study will review the basic operation of the ATJ and the spectrum analysis algorithm.

\section{ATJ Basics}
\label{s:ATJBasic}

\begin{figure}
\centering
\includegraphics[width=\columnwidth]{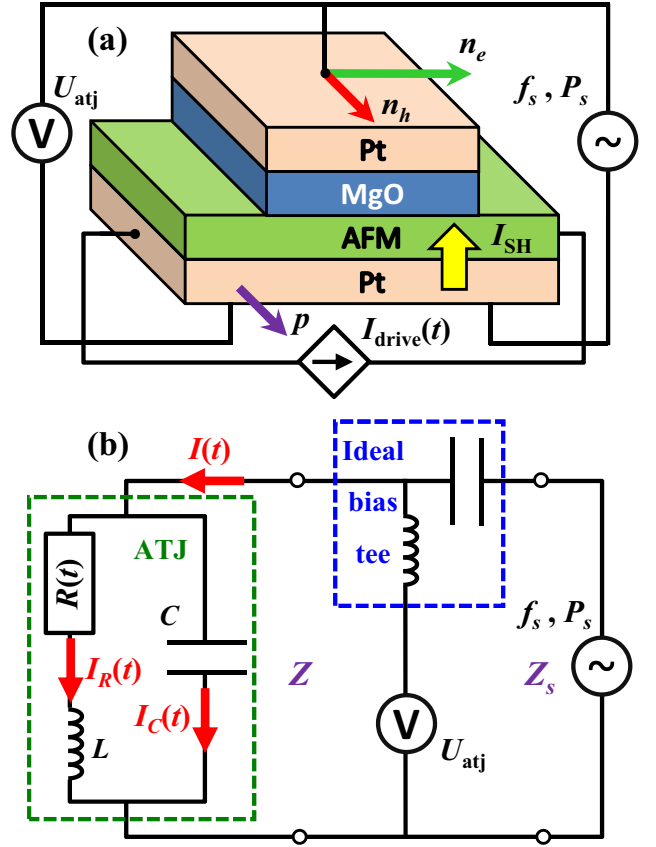}
\caption{
(a) Pt/AFM/MgO/Pt antiferromagnetic tunnel junction (ATJ).
The driving dc current $I_{\rm drive}(t)$ flowing in the bottom Pt layer of an ATJ generates the transverse spin current $I_{\rm SH}$ flowing into the AFM layer, which excites the TF rotation of magnetization of the AFM sublattices, and, consequently, the TF variations of the junction resistance $R(t)$. When a TF signal with power $P_s$ and frequency $f_s$ is supplied to the junction, its action results in the generation of voltage $U_{\rm atj}$ across the whole structure;
(b) A simplified equivalent electric scheme of the ATJ-based detector connected through an ideal bias tee to a source of an external TF signal.
}
\label{f:Model}
\end{figure}

The physical structure of an ATJ is shown in Fig. \ref{f:Model}(a) \cite{Sulymenko2018IEEEML}.
It consists of a four-layer structure with a lower platinum (Pt) layer, a conducting AFM layer, a magnesium oxide (MgO) layer that serves as a tunneling barrier, and an upper platinum layer.
Oscillations arise in the following manner.
First, the driving dc current $I_{\rm drive}$ flows through the bottom Pt layer, generating a transverse spin current $I_{\rm SH}$ due to the spin Hall effect \cite{Sinova2015RMP}.
This spin current penetrates the AFM/Pt interface and excites TF rotation of the AFM sublattices and the AFM Neel vector\cite{Khymyn2017SciRep}.
The frequency of the TF oscillations is directly dependent on the magnitude of $I_{\rm drive}$.
This means that the operating frequency of the ATJ can be dynamically tuned by simply changing the dc bias $I_{\rm drive}$.

The oscillations can be extracted from the AFM layer by tunneling anisotropic magnetoresistance (TAMR) \cite{Park2011NatMater}.
TAMR will be present when the inverse spin current in the AFM layer tunnels through the dielectric MgO layer barrier to the upper Pt layer \cite{Sulymenko2018IEEEML}.
Experimentally, TAMR electrical switching in an ATJ was observed to be dependent on the orientation of the AFM Neel vector \cite{Park2011NatMater, Wadley2016Sci, kriegner}.
Thus, when the AFM Neel vector rotates with a TF, the TAMR also oscillates with the same TF.
In this paper, we will treat the TAMR as a macroscopic oscillating resistance, which we call $R(t)$.


\section{Spectrum Analysis Algorithm}
\label{s:saa}

Spectrum analysis is performed with an algorithm that is similar to a previous study\cite{Louis2018APL}.
In that study, it was demonstrated that a magnetic tunnel junction can, in theory, perform spectrum analysis between 26 and 36 GHz in 10 ns, with a frequency resolution near the theoretical limit.
It is important to note that the algorithm presented in that study is not specific to any particular technology, and any auto-oscillator that can be linearly tuned through a scanning bandwidth can be used.
In this regard, because ATJs are in principle easily tunable through the TF range, they are a good candidate for TF spectrum analysis.

In this algorithm, $R(t)$ oscillates with a THz frequency $f_{\rm R}(t)$ that varies with time.
Specifically, when the bias dc current $I_{\rm drive}\rightarrow I_{\rm drive}(t)$ increases linearly with time, $f_{\rm R}(t)$ also increases linearly as:
\begin{equation}\label{eq:fr}f_{\rm R}(t) = f_0 + \rho t~.\end{equation}
In this equation, $f_0$ is an initial frequency, and $\rho$ is the rate of frequency change of the ATJ, which can be measured in units of GHz/ns or THz/ps.
Practically, $\rho$ is the scan rate at which spectrum analysis is performed through a particular bandwidth $\Delta f$.
When the TAMR effect is considered to be an oscillating resistance $R(t)$, it is modeled as:
\begin{subequations}
\label{eq:Basics}
\begin{equation}
\label{eq:BasicsRt}
   R(t) = R_0 + \Delta R \cos\left(2 \pi f_0 t + \pi \rho t^2 \right)
   \, ,
\end{equation}
where $R_0$ is the equilibrium resistance of the ATJ, and $\Delta R$ is the magnitude of the variation of the junction ac resistance.

Spectrum analysis will be performed on an external TF current, given by
\begin{equation}
\label{eq:BasicsIRt}
  I_R(t) = I_R \cos\left(2 \pi f_s t + \phi\right)
   \, ,
\end{equation}
where $I_R$ is the magnitude of the current flowing through the oscillating resistance of the ATJ, $f_s$ is a constant frequency that is within $\Delta f$, and $\phi$ is the initial phase of the external current.

When $I_R(t)$ flows through the ATJ, it will be multiplied by $R(t)$ to produce a voltage $U_{\rm atj}$.
This voltage will have three terms.
The first term, $R_0 I_R(t)$, will have a THz frequency.
The second and third terms will have frequencies $(f_R(t) + f_s)$ and $(f_R(t) - f_s)$, resulting from product of the cosines in (\ref{eq:BasicsRt}) and (\ref{eq:BasicsIRt}).
The $(f_R(t) + f_s)$ term will have a THz frequency, while the $(f_R(t) - f_s)$ term will at times have a low frequency.
When a low pass filter with an appropriately chosen cutoff frequency $f_c$ is applied to $U_{\rm atj}$, only the $(f_R(t) - f_s)$ term will remain. The output voltage of the filter is given by
\begin{equation}
\label{eq:BasicsUlpf}
   U_{\rm lpf}
   = \tfrac{1}{2} g(t)I_R \Delta R \cos(\theta(f_s,t)-\phi)
   \, ,
\end{equation}
where $\theta(f_s, t)=2\pi(f_0 - f_s)t + \pi \rho t^2$.
The function $g(t)$ is a dimensionless function that represents a low pass filter that behaves as follows.
If $\theta(f_s, t) < 2\pi f_c t$, then $U_{\rm lpf}$ will be in the pass band of the filter and thus $g(t)=1$.
If $\theta(f_s, t) \gg 2\pi f_c t$, then $U_{\rm lpf}$ will be in the stop band and $g(t)$ will be very small, $g(t)\ll 1$.
When $\theta(f_s, t) \gtrapprox  2\pi f_c t$, the filter is in the transition region and $g(t)$ will depend on the order and type of the low pass filter.

It is important emphasize that the signal $U_{\rm lpf}$, which has passed through a low pass filter, contains only frequencies that are below $f_c$.
For example, there will be times when the frequency of the external signal $f_s$ is identical to that of the ATJ.
At these times, when $f_s = f_0 +\tfrac{1}{2}\rho t$, for a brief moment in time $U_{\rm lpf}$ will be a simple dc signal, when $\theta(f_s, t)=0$.
Thus, despite the fact the both the external signal to be analyzed and the ATJ both have frequencies in the THz gap, when they are mixed $U_{\rm lpf}$ will have a frequency near dc.
Thus, it is possible to use any traditional spectrum analysis algorithm, for example, envelope detection of matched filtering, which is detailed as follows.

The next step in the spectrum analysis algorithm is to apply a matched filter to $U_{\rm lpf}$.
The previous study\cite{Louis2018APL} shows that the output spectrum can be obtained by
\begin{equation}
\label{eq:BasicsMatch}
   U_{\rm spec}=h(f_m,t)*U_{\rm lpf}
   \, ,
\end{equation}
\end{subequations}
where $U_{\rm spec}$ is the analyzed spectrum, $*$ is the symbol for convolution, and the matched filter is given by $h(f_m,t)=\exp\left[i\theta(f_m,t) \right]$, where $f_m$ is an arbitrary frequency in the interval of spectrum analysis $\Delta f$.
The matched filter serves several purposes, including improving the SNR and removing the dependance on the phase difference $\phi$ between $R(t)$ and $I_R(t)$. When $U_{\rm lpf}$ has been filtered by a matched filter, it contains the frequency spectrum of $I_{R}(t)$ encoded in time.

There are several important characteristics to note concerning this algorithm.
Firstly, the magnitude of $U_{\rm spec}$ does not depend on the magnitude of $U_{\rm lpf}$; for effective spectrum analysis, all that is required is that the external signal be within 3 dB of the Johnson-Nyquist thermal noise floor.
For this reason, much of the following analysis in this manuscript will focus on $U_{\rm lpf}$.
Secondly, this algorithm can faithfully analyze the spectrum for complicated signals, for example, when $I_R(t)$ has multiple frequencies with multiple amplitudes, spectrum analysis can be performed for all frequencies within $\Delta f$.
However, for simplicity, in most of this manuscript, $I_R(t)$ will be assumed to have only a single frequency; it will be a monochromatic signal.
Lastly, although we are detecting signals in the THz frequency region, $U_{\rm lpf}$ will have a frequency low enough to allow the use of standard analog to digital converters, which can simplify signal processing by allowing $h(f_m,t)$ to be applied in the digital domain.

It is important to mention the frequency resolution that can be attained using this algorithm.
The measure of the minimum separation required for a spectrum analyzer to distinguish two neighboring signals is commonly called resolution bandwidth (RBW).
The RBW theoretical limit for swept tuned spectrum analyzers can be given by the equation\cite{gabor}:
\begin{equation}
\label{rbw}
	\textrm{RBW}\approx \frac{\rho}{f_c}
	\, .
\end{equation}
Numerical simulations performed in \cite{Louis2018APL} showed that this algorithm performs according to (\ref{rbw}).

The cornerstone of this algorithm is that for this particular matched filter to function, the ATJ must operate with a frequency that increases linearly according to (\ref{eq:fr}).
The next section will discuss the ability of the ATJ to tune linearly through a finite frequency range.

\section{ATJ Dynamics}
\label{s:ATJD}

An ATJ can be used for this spectrum analysis algorithm because its oscillation frequency can be dynamically tuned in a simple manner; all that is required is a change in the magnitude of the bias current $I_{\rm drive}$.
In this section it will be demonstrated that the frequency of an ATJ can be dynamically tuned so that the frequency increases linearly with time.
A simple way to demonstrate that the ATJ is capable of being tuned according to (\ref{eq:fr}) with a constant scan rate $\rho$ is to perform numerical simulations of the magnetization dynamics of the ATJ.

The dynamics of the AFM layer can be modeled with a pair of coupled Landau-Lifshitz-Gilbert-Slonczewski equations\cite{Khymyn2017SciRep}
\begin{subequations}
\label{eq:llgs}
\begin{equation}
\label{eq:llgs1}
  \frac{d\bm{m}_1}{dt}=\gamma \bm{B}_1 \times \bm{m}_1 +\alpha_{\rm eff}\bigg[\bm{m}_1\times\frac{d\bm{m}_1}{dt}\bigg]+ j \sigma[\bm{m}_1\times[\bm{m}_1\times \bm{p}]]
   \, ,
\end{equation}
\begin{equation}
\label{eq:llgs2}
   \frac{d\bm{m}_2}{dt}=\gamma \bm{B}_2 \times \bm{m}_2 +\alpha_{\rm eff}\bigg[\bm{m}_2\times\frac{d\bm{m}_2}{dt}\bigg]+ j \sigma[\bm{m}_2\times[\bm{m}_2\times \bm{p}]]
   \, .
\end{equation}
\end{subequations}
Where $\bm{m}_1$ and $\bm{m}_2$ are the normalized unit-length magnetization for the two sublattices in the AFM material in the macrospin approximation. In these equations, $\bm{B}_1$ and $\bm{B}_2$ are the effective magnetic fields acting on the sublattices $\bm{m}_1$ and $\bm{m}_2$:
\begin{subequations}
\label{eq:effH}
\begin{equation}
\label{eq:effH1}
  \bm{B}_1=-\tfrac{1}{2}B_{ex} \bm{m}_2 - B_h \bm{n}_h(\bm{n}_h \cdot \bm{m}_1) +  B_e \bm{n}_e(\bm{n}_e \cdot \bm{m}_1)
   \, ,
\end{equation}
\begin{equation}
\label{eq:effH2}
   \bm{B}_2=-\tfrac{1}{2}B_{ex} \bm{m}_1 - B_h \bm{n}_h(\bm{n}_h \cdot \bm{m}_2) +  B_e \bm{n}_e(\bm{n}_e \cdot \bm{m}_2)
   \, .
\end{equation}
\end{subequations}

As this study is focused on qualitative behavior of AFM materials, we have adapted the simulation parameters used in \cite{Khymyn2017SciRep} for an easy plane conducting AFM material.
When performing simulations according to equations (\ref{eq:llgs}) and (\ref{eq:effH}), we chose physical parameters as follows.
The gyromagnetic ratio is $\gamma = 2\pi 28$~GHz/T, $\alpha_{\rm eff}=$0.01 is the effective Gilbert damping parameter, and $j$ is the electric current density in units of A/cm$^2$.
The unit vector along the spin current polarization $\bm{p}$ is directed as shown in in Fig. \ref{f:Model}(a)
The spin torque coefficient is given by $\sigma= \gamma e \frac{ \theta_{sh} \lambda \rho_0 g_r}{ M_s d_{\rm AFM}}\tanh\frac{d_{\rm Pt}}{2\lambda}$, where $e=1.602 \times 10^{-19} $~C is the fundamental electric charge.
For the lower Pt layer, $\theta_{sh}=0.1$ is the spin Hall angle, $\lambda=7.3$ nm is the spin diffusion length in Pt, and $\rho_0=4.8\times10^{-7} ~\Omega\cdot \rm{m}$ is the electrical resistivity\cite{wang2014}.
The thickness of the Pt layer is assumed to be $d_{\rm Pt}=20$ nm.
The parameter $g_r= 7.0\times10^{18} m^{-2}$ is used for the spin-mixing conductance at the Pt-AFM interface\cite{cheng2014}, $M_s= 350$ kA/m is the magnetic saturation of one AFM sublattice, $d_{\rm AFM}=1$ nm is the thickness of the AFM layer.
The exchange frequency is chosen to be $\omega_{ex}=\gamma H_{ex}=2\pi\cdot 60$ THz, $H_h$ is the hard-axis anisotropy field with $\omega_h= \gamma H_h = 2\pi \cdot 30$ GHz, and $H_e$ is the easy axis anisotropy field with $\omega_e= \gamma H_e = 2\pi \cdot 0.1$ GHz.

\begin{figure}
\centering
\includegraphics{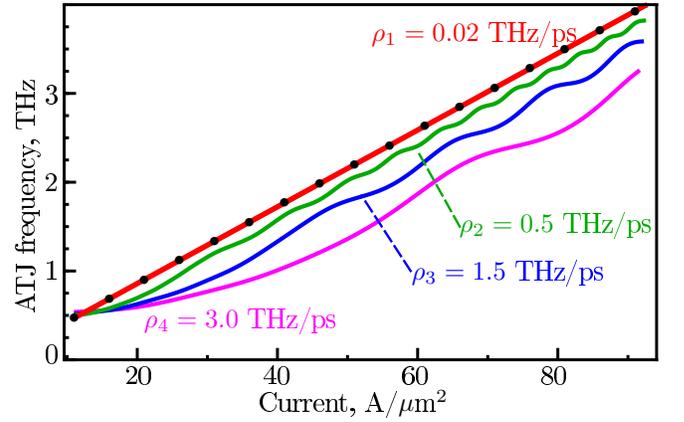}
\caption
{
ATJ dynamic response to ramped current. For each $\rho$, the ATJ runs for 50 ps at 0.5 THz to reach normal operation. Then, the current is ramped to cause linear frequency increase at the following rates: $\rho_1=0.02$ THz/ps, $\rho_2=0.5 $ THz/ps, $\rho_3= 1.5$ THz/ps, and $\rho_4=3.0$ THz/ps.
}
\label{f:atjDyn}
\end{figure}

Results of the simulation for several linearly increasing bias currents are shown in Fig. \ref{f:atjDyn}.
In this figure, the black dotted line shows the static relationship between $I_{\rm drive}$ and $f_R$.
The red line shows the frequency output when the ATJ is tuned with a scan rate of $\rho_1=0.02$~THz/ps.
It is evident that at this scan rate the ATJ tunes in a quasi-static manner, and that theoretically, an ATJ can be tuned linearly with time.
This is also true for slower scan rates.
At a faster scan rate of  $\rho_2=0.5 $~THz/ps, shown by a green line, there is a slight offset from quasi-static, and a slight ripple.
This ripple is arises due to the inertial dynamics of the AFM sublattices, and is related to the transient forced oscillations of the system.
The presence of these ripples points to a physical limit for a maximum $\rho$ where the ATJ ramps linearly with time.
At even faster scan rates, $\rho_3= 1.5$~THz/ps (blue) and $\rho_4=3.0$~THz/ps (magenta), the offset from linear dependence increases, as does the amplitude of the ripple.
In this study, we will perform spectrum analysis at a scan rate of $\rho=$~10~GHz/ns, in the quasi-static regime, where $f_{\rm R}(t)$ increases according to (\ref{eq:fr}).

We have thus shown via numerical simulation that an ATJ is capable of being tuned in a dynamic manner at a rate that allows the frequency to increase linearly with time.
This means that an ATJ can be used with the matching filter in (\ref{eq:BasicsMatch}), allowing spectrum analysis to be performed.


\section{Electrical Model of the ATJ}
\label{s:Model}

The previous sections described the spectrum analysis algorithm and the dynamics of the ATJ.
This section will describe the impact that parasitic capacitance $C$ and parasitic inductance $L$ will have on $U_{\rm lpf}$.
$L$ and $C$ must be considered in this system because at THz frequencies, matching losses cannot be eliminated by reducing the fabricated circuit size.
This section will present the equivalent circuit to the ATJ, characterize matching losses, as well as describe the electrical parameters of the ATJ.\

In this section, it is prudent to treat the ATJ as a detector of single frequencies; specifically, by setting $\rho=0$, $f_0=f_s$, and $\phi=0$.
When operating with these parameters, the ATJ will oscillate with exactly the same frequency and phase as $I_{\rm R}$. With this condition, $U_{\rm lpf}$ from (\ref{eq:BasicsUlpf}) reduces to a dc voltage, which is stated here for clarity:

\begin{equation}
\label{eq:Basics-Udc}
   U_{\rm dc}
   = \tfrac{1}{2} I_R \Delta R
   \, .
\end{equation}

\subsection{Circuit and basic equations}

Through circuit analysis, the matching loss for the external signal when interfacing with the ATJ can be analyzed\cite{Sulymenko2018IEEEML}.
The ATJ shown in Fig. \ref{f:Model}(a) can be modeled with the equivalent electric circuit shown in Fig. \ref{f:Model}(b).
This electric scheme consists of the three parts:
(i) the equivalent circuit of an ATJ, which is bounded by a dashed green line and characterized by the frequency-dependent impedance
$Z \equiv Z(f_s)$;
(ii) the external circuit that has an impedance $Z_s$;
and (iii) a bias tee, which is bounded by a dashed blue line and provides coupling between the ATJ and an external circuit.

It should be stressed that this circuit model does not include low-frequency support elements, such as a source for drive current $I_{\rm drive}$ and matching circuits.
Also, for simplicity in this subsection the bias tee is considered to be an ideal coupling element, which perfectly separates low frequency and THz signals in the circuit and does not influence the device performance.

The electric scheme of the ATJ includes one circuit branch with the oscillating resistance $R(t)$ characterized by the equilibrium resistance $R_0$ and the inductance $L$ of an ATJ having the impedance $Z_L = \ii \omega_s L$ (here $\omega_s = 2 \pi f_s$ is the angular frequency of the input signal, and $\ii = \sqrt{-1}$).
This is connected in parallel to the other branch with the junction capacitance $C$ described by the impedance $Z_C = 1/\ii \omega_s C$.
The frequency-dependent complex impedance of the ATJ, which is $Z = (R_0 + Z_L) Z_C / (R_0 + Z_L + Z_C)$, can be written in the form:
\begin{subequations}
\label{eq:Z}
\begin{eqnarray}
   Z
   &=&
   R + \ii X
   \, ,
   \\
   R
   &=&
   {\rm Re}\{Z\}
   =
   \frac{R_0}{q}
   \, ,
   \\
   X
   &=&
   {\rm Im}\{Z\}
   =
   \frac{\omega_s L (1 - \xi) - R_0 \beta}{q}
   \, .
\end{eqnarray}
\end{subequations}
Here we introduce two dimensionless parameters:
$\xi = \omega^2_s L C$, which describes resonance properties of the ATJ, and
$\beta = \omega_s R_0 C$, which characterizes inertial properties of the ATJ,
and the ansatz $q = (1 - \xi)^2 + \beta^2$.

The input ac current $I(t)$ can be defined as $I(t) = I \cos(\omega_s t)$, where $I$ is the current magnitude.
Then the complex amplitudes $\hat{I}_R$, $\hat{I}_C$ of ac currents $I_R(t)$, $I_C(t)$, respectively, in the circuit shown in Fig.~\ref{f:Model}(b) are governed by  Kirchhoff's laws:
\begin{subequations}
\label{eq:System}
\begin{eqnarray}
   \hat{I}_R + \hat{I}_C &=& I
   \, ,
   \\
   \hat{I}_R (R_0 + \ii \omega_s L) &=& \hat{I}_C \frac{1}{\ii \omega_s C}
   \, .
\end{eqnarray}
\end{subequations}
The solution of this system of equations can be written in the form:
\begin{subequations}
\label{eq:Sol}
\begin{eqnarray}
\label{eq:Sol-IR}
   \hat{I}_R   &=&   I \frac{1}{1 - \xi + \ii \beta}
   \, ,
   \\
\label{eq:Sol-IC}
   \hat{I}_C   &=&   I \frac{- \xi + \ii \beta}{1 - \xi + \ii \beta}
   \, .
\end{eqnarray}
\end{subequations}
Thus from the real part of (\ref{eq:Sol-IR}) the output voltage $U_{\rm dc}$, given by Eq.~(\ref{eq:Basics-Udc}), can be rewritten as
\begin{equation}
\label{eq:UdcII}
   U_{\rm dc} = \tfrac{1}{2}\bigg[\frac{1 - \xi}{q} I\bigg]\Delta R
   \, .
\end{equation}

The magnitude of input ac current $I$ can be determined from the equation $P_s (1 - |\Gamma|^2) = 0.5 I^2 R$, which describes the transfer of average input signal power $P_s$ from an external TF electrodynamic system to an ATJ, where $\Gamma = (Z-Z_s)/(Z+Z_s)$ is the complex reflection coefficient.
The real part of the impedance $Z_s$ of an external circuit, $R_s = {\rm Re}\{Z_s\}$, usually can be considered as a constant value within a rather narrow frequency range, while the imaginary part of the impedance, $X_s = {\rm Im}\{Z_s\}$, changes with the frequency $f_s$, but can be controlled using impedance matching techniques \cite{Mcw}. In the following for simplicity we assume that $R_s$ is constant and $X_s$ = 0.

In this case an expression for the output dc voltage $U_{\rm dc}$ can be written in the form:
\begin{equation}
\label{eq:UdcPs}
   U_{\rm dc}   =   \frac{1 - \xi}{q}		\sqrt{\frac{2 P_s}{R} }	\sqrt{  \frac{R R_s}{(R + R_s)^2 + X^2}   }   \Delta R
   \, .
\end{equation}

$U_{\rm dc}$ strongly depends on the matching coefficient $(R R_s)/[(R + R_s)^2 + X^2]$ under the square root in (\ref{eq:UdcPs}).
Hence, the ATJ can have good performance only if the active junction impedance $R$ and the active impedance $R_s$ of an external circuit are almost equal ($R \approx R_s$).
Usually, the active impedance $R_s$ is considered to be a fixed parameter, defined by the properties of the external TF electrodynamic system.
In the case of ideal matching, when $R = R_s$, the matching coefficient $(R R_s)/[(R + R_s)^2 + X^2]$ is equal to $R_s^2/[4 R_s^2 + X^2]$, which gives an output dc voltage for a perfectly matched detector:
\begin{equation}
U_{\rm dc} =  \frac{1 - \xi}{q}	 \sqrt{\frac{2 P_s}{R}} \sqrt{  \frac{R_s^2}{4R_s^2 + X^2}   }\Delta R,
\end{equation}
or with Eq. (\ref{eq:Z}),
\begin{equation}
\label{eq:UdcMatch}
   U_{\rm max}    =    \frac{1 - \xi}{\sqrt{q}} \sqrt{\frac{2 P_s}{R_0}}\sqrt{\frac{1}{4+\zeta^2}}\Delta R
   \, ,
\end{equation}
where $\zeta=X/R_s$.

To reach the optimal condition $R \approx R_s$, one can vary the cross-sectional area $S$ of the junction and the thickness $d$ of the MgO tunneling barrier, as will be presented in the discussion.

\subsection{Electrical parameters of the ATJ}
\label{electricalParamters}
Our electrical model of the ATJ can be characterized by three intrinsic parameters: the junction equilibrium resistance $R_0$, the inductance $L$ and the capacitance $C$.

Using the approach introduced in \cite{Sulymenko2018IEEEML}, we assume that the equilibrium resistance $R_0$ depends on the thickness of the MgO barrier $d$, the junction cross-sectional area $S$, the ATJ effective resistance-area product ${\rm RA}(0)$ (introduced for a ``zero-thickness'' MgO barrier), and the intrinsic MgO tunneling barrier parameter $\kappa$ \cite{Hayakawa2005JpnJAP,Yuasa2007JPhysD}.
This can be written in the form:
\begin{equation}
\label{eq:R0}
   R_0 \equiv R_0(S, d)
   =
   \frac{{\rm RA}(0)\exp\left(\kappa d\right)}{S}
   \, .
\end{equation}
Note that the chosen dependence of the resistance-area product of an ATJ on the thickness $d$ of the tunneling barrier, ${\rm RA}(d)={\rm RA}(0)\exp\left(\kappa d\right)$, have the same form as that for a conventional ferromagnetic tunnel junction \cite{Hayakawa2005JpnJAP,Yuasa2007JPhysD}.

The ac resistance variations $\Delta R$ of an ATJ can be evaluated using the TAMR ratio $\eta$ as
\begin{equation}
\label{eq:DeltaR}
   \Delta R
   =
   \frac{\eta}{2 + \eta} R_0
   \, .
\end{equation}
The TAMR ratio $\eta$ can be calculated for given $\Delta R$ as
$\eta = 2 \Delta R / (R_0-\Delta R)$.

The capacitance of an ATJ can be estimatd as the capacitance of a square parallel-plate capacitor with a plate size $a = \sqrt{S}$ and the distance $d$ between the plates:
$C = \varepsilon \varepsilon_0 S / d$ ($\varepsilon$ is the MgO relative permittivity \cite{Raj2010MgO}, and $\varepsilon_0$ is the vacuum permittivity).
The inductance of an ATJ can be evaluated as $L = \mu_0 d$, where $\mu_0$ is the vacuum permeability.

Values for these parameters can be estimated using experimental parameters as presented in Ref \cite{Park2011NatMater}:
equilibrium resistance $R_0 = 55.0$~k$\Omega$,
TAMR ratio $\eta = 1.3$.
The value of the intrinsic MgO barrier parameter can be estimated as $\kappa \approx 5.8$~nm${}^{-1}$ if we assume that the dependence of the junction resistance $R_0$ on the tunneling barrier thickness $d$ for an ATJ is similar to that for a ferromagnetic junction \cite{Hayakawa2005JpnJAP,Yuasa2007JPhysD}.
Using these values, the effective resistance-area product is ${\rm RA}(0) \approx 0.14 \ \Omega \cdot \mu {\rm m}^2$ and the magnitude of the ac resistance is $\Delta R \approx 21.7$~k$\Omega$.
Finally, $\varepsilon = 9.8$ is a reasonable estimate for the relative permittivity of the MgO barrier \cite{Raj2010MgO}.

For the ATJ presented in Ref \cite{Park2011NatMater}, the junction cross-sectional area was $S = 5$~$\mu$m${}^2$, and the thickness of the MgO barrier was $d = 2.5$~nm.
Using these values, we can estimate a junction capacitance of $C = \varepsilon \varepsilon_0 S/d = 170$~fF and an internal inductance of $L = \mu_0 d = 3.1$~fH for the ATJ in that experiment.
Note that for these parameters, $f_{resonance}=1/(2\pi\sqrt{LC}) \approx 7$ THz.
As this frequency is near our frequency of interest, it must be taken into account.
As will be explained in the discussion, $S$ and $d$ will be varied to adjust $R$, $C$, and $L$ and thus optimize ATJ performance.

Finally the value of the output voltage $U_{\rm max}$ cannot exceed the breakdown voltage $U_{\rm b} \simeq E_{\rm b} d$, where $E_{\rm b}$ is the breakdown electric field for the tunneling barrier.
$E_{\rm b} = 0.4 - 0.6$~V/nm for a MgO thin film \cite{Raj2010MgO}.

\section{Results and discussion}
\label{s:Results}

This section begins with a discussion about optimizing the ATJ for operation in the static regime, then briefly discuss the ATJ as a detector of single frequencies. Then, numerical simulation of spectrum analysis in the THz gap with an ATJ will be presented.

\subsection{Results and discussion for static regime}

Here, we will discuss the optimization of the ATJ design to allow $U_{\rm dc}$ to be greater than 1 mV. This will be achieved by optimizing parameters $d$ and $S$ to maximize the transfer of input signal power to the ATJ.

We begin by discussing an ATJ with parameters as in the previous experimental work \cite{Park2011NatMater}.
Specifically, if the active impedance $R_s$ of the external circuit has a value of $R_s = 50 ~ \Omega$, which is typical for microwave and terahertz electronics, and the input signal power is $P_s = 1$~$\mu$W, the maximum TF output will be $U_{\rm dc} \sim 10^{-4}$~mV at $f_s \sim 0.1$~THz.
This small voltage is primarily due to the large values of the junction resistance $R_0$ and capacitance $C$, which causes poor impedance matching and reduction of ac power transferred to the junction.
We will show that by optimizing the dimensions of the dielectric layer in the ATJ, the output voltage can be substantially increased.

\begin{figure}
\centering
\includegraphics{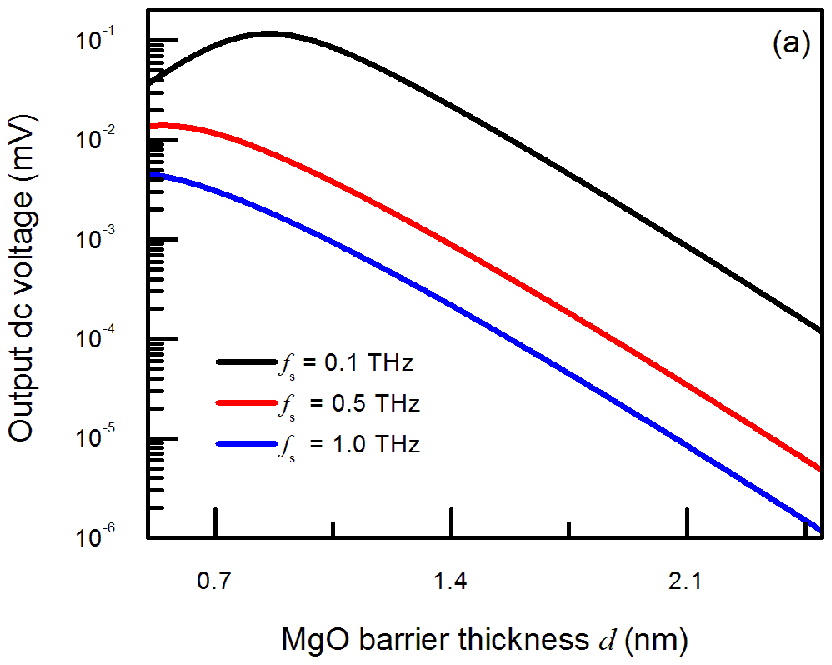}
\includegraphics{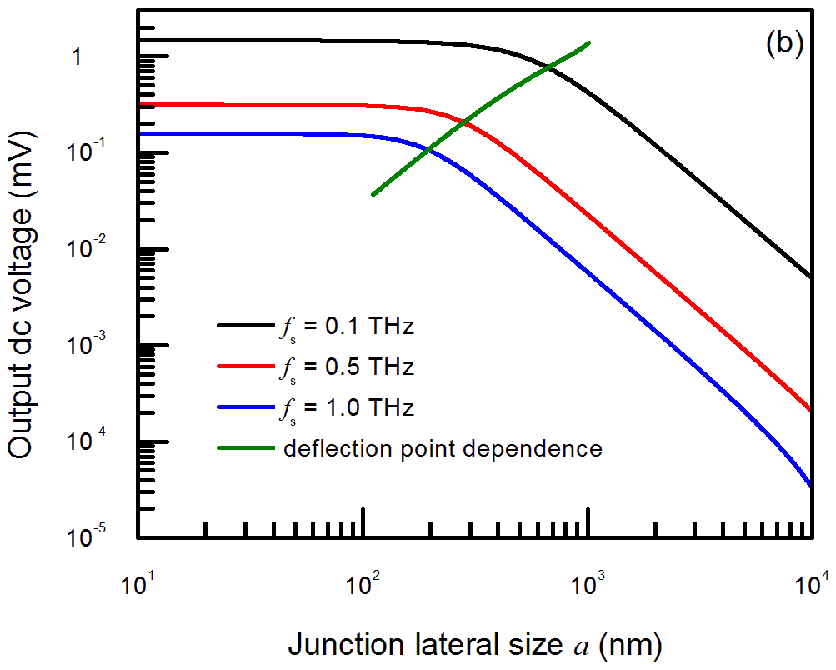}
\caption
{
Dependence of the output dc voltage $U_{\rm dc}$ at input power of 1~$\mu$W on (a) MgO barrier thickness $d$ and (b) lateral size of the ATJ square-cross-section $a = \sqrt{S}$.
Calculations were performed using (\protect\ref{eq:UdcPs}) and (\protect\ref{eq:Z}), (\protect\ref{eq:R0}), (\protect\ref{eq:DeltaR}) for an ATJ with parameters similar to those observed in experiment (see Section \protect\ref{electricalParamters} for details) operating at the frequencies $f_s = 0.1$~THz (black solid line), $f_s = 0.5$~THz (red line), and $f_s = 1.0$~THz (blue line).
In (b), the optimal thickness of the MgO barrier $d_{\rm opt} = 1.0$~nm was used. Green line shows curve that represents dependence of output dc voltage at deflection point for different frequencies.
}
\label{f:UdcdS}
\end{figure}

This ATJ can be optimized for optimal performance by choosing appropriate geometrical junction parameters and thus the ATJ equilibrium resistance $R_0 \sim \exp (\kappa d)/S$, junction inductance $L \sim d$, and the junction capacitance $C \sim S/d$.
In Fig.~\ref{f:UdcdS}, we examine the relationship between $U_{\rm dc}$, $S$, and $d$ in order to optimize ATJ performance.

Fig.~\ref{f:UdcdS}(a) shows that in general, $U_{\rm dc}$ increases monotonically with the decrease of $d$, the MgO layer thickness.
$U_{\rm dc}$ can have a maximum value at low frequencies of $d$.
For example, when $f_s = 0.1$~THz, there is a maximum at $d \approx 0.8$~nm.
This occurs when the active junction impedance $R$ becomes comparable to the external active impedance $R_s$.
For values of $d\lesssim 1$ nm, $U_{\rm dc}$ plateaus or decreases instead of increasing monotonically.
This plateau or decrease is due to the increase of the impedance mismatch and junction capacitance. 
Taking this into account, in the following simulations we consider an ATJ having a MgO barrier thickness of $d_{\rm opt} = 1.0$~nm, which is a common value for existing tunnel junctions and can be readily fabricated \cite{Hayakawa2005JpnJAP,Yuasa2007JPhysD,Raj2010MgO}.

An additional improvement of the TF signal detector performance can be achieved by varying the junction cross sectional area $S$.
This will be considered in Fig.~\ref{f:UdcdS}(b).
For simplicity, we assume a square-shape junction with a single effective lateral junction size $a = \sqrt{S}$.
As one can see that for different frequencies, the value of $U_{\rm dc}$ is constant for lower values of  lateral junction size $a$.
At these low $a$ values, the performance of the signal detector is hindered mainly due to the resistance mismatch effect.
For higher values of $a$, the value of $U_{\rm dc}$  decreases.
For high $a$ values, the device efficiency is mainly reduced due to the influence of the large junction capacitance.
At junction sizes where $R-R_s \approx X$, as shown by the green dotted line in the figure, the influence of junction capacitance and resistance mismatch have similar levels of influence.
While there is no optimal value for $a$, it is evident that lower values, below a certain point, lead to improved behavior. Also, one should consider, that the smaller the size, the harder to fabricate the junction.
Therefore, we have chosen a junction size of $a_{opt}= 300$~nm.

\begin{figure}
\centering
\includegraphics{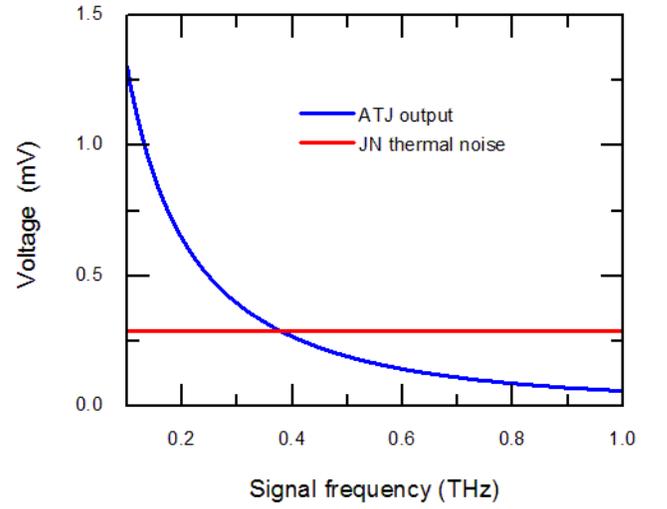}
\caption
{
Frequency dependence of the output dc \mbox{voltage}~$U_{\rm dc}$ (blue line) and the Johnson-Nyquist thermal noise floor rms voltage (red line) calculated from (\protect\ref{eq:UdcPs}) using typical ATJ parameters stated in Section \protect\ref{electricalParamters}, and parameters $d~=~d_{\rm opt}~=~1.0$~nm and $a = 300$~nm at the temperature $T=293$~K under input power of 1~$\mu$W.
}
\label{f:UdcF}
\end{figure}

Fig.~\ref{f:UdcF} shows the output dc voltage $U_{\rm dc}$ calculated numerically from (\ref{eq:UdcPs}) with new spacial dimensions of $d_{\rm opt} = 1.0$~nm and $a_{\rm opt} = 300$~nm.
This graph demonstrates that with optimized physical dimensions, the ATJ is capable of producing a strong dc voltage output with a reasonably sized input signal.
This figure also demonstrates that $U_{\rm dc}$ in the frequency range $0.1-1.0$~THz from an optimized AFM-based detector connected to a standard impedance load, is comparable to, or may even exceed, the dc voltage extracted via an inverse spin Hall effect from the detector based on a bi-anisotropic NiO/Pt structure \cite{Khymyn2017AIPAdv}.
Also in contrast to the detector based on bi-anisotropic NiO/Pt spin Hall oscillator \cite{Khymyn2017AIPAdv}, the ATJ in this system does not require any special conditions for the AFM layer, therefore the experimental realization seems to be relatively straightforward.

With these new dimensions, the device parameters will be $C_{\rm opt}=7.81$~fF, and $L_{\rm opt}=1.26$~fH.
It is noteworthy that for both the optimized parameters and the ATJ presented in Ref \cite{Park2011NatMater}, that $\xi \ll 1$ for signal frequencies $f_s \le 2$~THz.
This means that the nonlinear coefficient $(1-\xi)/\sqrt{q}$ in equations (\ref{eq:UdcII}) through (\ref{eq:UdcMatch}) can be approximated by $(1-\xi)/\sqrt{q}\approx 1/\sqrt{1 + \beta^2} \approx 1/\sqrt{1 + \omega_s^2 R_0^2 C^2}$.
This behavior is clearly visible in Fig.~\ref{f:UdcF}.

Before concluding this section, we want to note that according to Eqs. (\ref{eq:UdcPs}) and (\ref{eq:DeltaR}), the output dc voltage of the detector
$U_{\rm dc} \sim \Delta R \sim \eta/(2+\eta)$ increases with the increase of the TAMR ratio $\eta$.
Although in this paper we consider the TAMR ratio $\eta$ as a fixed parameter that is defined experimentally, in practice $\eta$ can be tuned in several ways, for example by adding a bias dc current through the junction, adding a bias dc magnetic field, or by changing the operating temperature of the ATJ.
However, the influence of the temperature on $\eta$ and $U_{\rm dc}$ in an ATJ-based detector could be substantially nonlinear, similar to the behavior of conventional ferromagnetic tunnel junctions \cite{Shang1998PRB,Prokopenko2012IEEETM}.

It is notable that with this analysis, an ATJ operating at a single frequency functions as a THz detector\cite{artemchuk}.
For such a detector it is reasonable to assume $\rho=0$ and $f_0=f_s$, however, in a detection application, generally the phase is unknown and thus one must use the detector with $\phi \ne 0$.
To prevent full attenuation of $U_{\rm lpf}$ in cases where $\theta(f_s,t) + \phi \approx \tfrac{\pi}{2}$, as calculated in Eq. (\ref{eq:BasicsUlpf}), signal averaging could be used to remove the phase dependance.
The general principles of operation ATJ based TF detector is similar to conventional spintronic detectors \cite{Tulapurkar2005Nature,Prokopenko2013Book,Prokopenko2015LTP} based on ferromagnetic materials.
However, there is one important difference: an ATJ-based detector can detect signals with substantially larger frequencies ($f_s \sim 0.1-10$ THz) than ferromagnetic detectors.

Thus concluding this section, we have shown that with the optimization of the thickness $d$ and area $S$ of the MgO layer, an ATJ can be designed to match an external signal and thus produce a strong dc voltage.
Specifically, there is an optimal $d$ to improve ATJ sensitivity, and decreasing $S$ also improves sensitivity.
We have also shown that the nonlinear coefficient $(1-\xi)/\sqrt{q}$ causes the sensitivity to decrease with increasing frequency.


\subsection{THz Spectrum Analysis with an ATJ}

\begin{figure}
\centering
\includegraphics{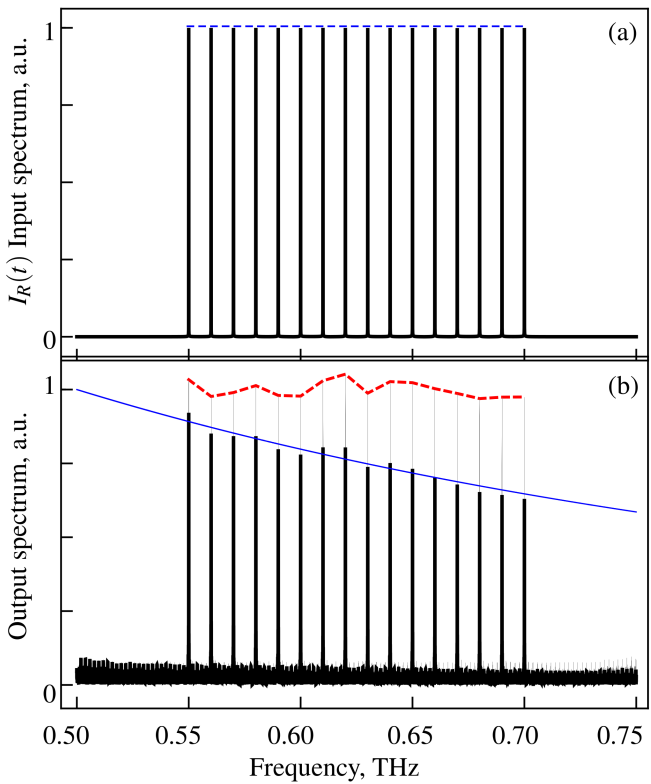}
\caption
{
Simulation of spectrum analysis by an ATJ.
(a) Input spectrum, shown by black lines, with a signal at every 10 GHz between 0.55 THz and 0.75 THz.
Blue dashed line shows envelope of input spectrum.
(b) Output spectrum for a scan from 0.5 to 0.75 THz, with a scan time of 25 ns.
The scan rate was 250 GHz/25 ns $\approx$ 10 GHz/ns.
Black curve shows $U_{\rm spec}$, including parasitic impedance.
Thin blue line represents (\protect\ref{f:UdcF}).
Red dashed line shows the envelope of $U_{\rm lpf}$ without parasitic impedance.
}
\label{spectrum}
\end{figure}

Thus far in this article, several things have been established.
First, it was demonstrated that an ATJ can produce a tunneling anisotropic magnetoresistance $R(t)$ that oscillates with a frequency that can be dynamically tuned over a wide frequency bandwidth in the THz gap.
Then a circuit model was presented that describes how the ATJ voltage output $U_{\rm lpf}$ depends on spurious capacitance $C$ and inductance $L$.
After this, the physical dimensions of the ATJ were optimized to improve the output voltage.
With these tasks complete, we now have the ingredients required to simulate spectrum analysis.

Spectrum analysis will be simulated in the bandwidth from 0.5~THz to 0.75~THz.
This bandwidth, $\Delta f=250$~GHz,  will be scanned at a rate of $\rho=10$~GHz/ns.
At this rate, the entire bandwidth can be scanned in 25~ns.
The ATJ as simulated will have parameters identical to the simulation used to produce Fig. \ref{f:atjDyn}, and the physical dimensions used to produce Fig. \ref{f:UdcF}.
Additionally, the low pass filter was simulated with a cutoff frequency $f_c=25$~GHz.
This $f_c$ was chosen so that the signal could, in principle, be digitized with a commercially available analog to digital converter.

To demonstrate the viability of the system, the analysis of a somewhat complicated spectrum will be simulated.
The input current $I_R(t)$ has a spectrum as shown in Fig. \ref{spectrum}(a), with signals every 10 GHz from 0.55 THz to 0.70 THz, each with a power of 1~$\mu$W.
The envelope of input spectrum is given by the blue dashed line.
We have assumed that these signals were generated with a high-Q AFM generator, and have a very low linewidth\cite{Sulymenko2018IEEEML}.
After following the algorithm presented in section \ref{s:saa}, an output spectrum was produced via simulation as shown in Fig. \ref{spectrum}(b).

At this power level, for frequencies $>0.4$~THz, the Johnson-Nyquist thermal noise floor rms voltage $U_{\rm jn}$ is larger than the ATJ output voltage $U_{\rm lpf}$.
Specifically, with the simulation parameters, $U_{\rm jn}=\sqrt{4 k_B T R_0 f_c}=0.29$~mV, where $k_B$ is the Boltzmann constant and the temperature is $T=293$ K.
A red curve in Fig. \ref{f:UdcF} shows $U_{\rm jn}$ in comparison to $U_{\rm dc}$. Despite the fact the thermal noise is larger than the input signal, the matched filter can produce an effective output.
This improvement of the SNR is a result of the matched filter, which can make the effective minimum detectable signal (MDS) is 3 dB below $U_{\rm jn}$.

The black curve in Fig. \ref{spectrum}b shows the output spectrum $U_{\rm spec}$, including the effects of parasitic impedance.
It is evident that the output curve has a spike that corresponds with every input frequency.
While in general the amplitude of the spikes is linear with $I_R \Delta R$, the amplitude of the $U_{\rm spec}$ has an offset that depends on both frequency and phase mismatch.
$U_{\rm spec}$ follows the thin blue line, which decreases with frequency according to $1/\sqrt{1+\omega_s^2 R_o^2 C^2}$ as in Fig. (\ref{f:UdcF}), as expected.
There is also a minor phase dependent variation in the amplitude of the output spectrum, as was discussed in \cite{Louis2018APL}.
The red dashed curve shows the envelope of $U_{\rm spec}$, while ignoring the effects of parasitic impedance.
This amplitude variation is expected, and can be removed by signal averaging or other means.
The relative error of these amplitude variations is about 2\%.
The bottom portion of the black curve shows interference that is the result of incoherent mixing from the matched filter.
The amplitude of this curve scales with signal power, and can limit dynamic range.
We wish to note that both types of amplitude variations can be easily normalized, as the signal processing occurs at low frequency according to (\ref{eq:BasicsUlpf}) and thus can occur in the digital domain.

Concerning frequency sensitivity, by using this algorithm, this simulated system was able to determine the frequency of the input spectrum with high precision and high accuracy.
Specifically, the peak of each spike in Fig. \ref{spectrum}(b) is within 5~MHz of the input frequency, with a relative error of $<10^{-4}\%$.

The lower end of the dynamic range is $>3$~dB below the Johnson-Nyquist thermal noise floor rms voltage $U_{\rm jn}$, as mentioned above.
The upper end of the dynamic range for the simulated parameters is determined by the breakdown voltage $U_{\rm b}$ as described at the end of section \ref{electricalParamters}.
With the dielectric MgO layer as simulated, $U_{\rm b}\approx 0.5$~V.
With this value, the total dynamic range for the ATJ is $\approx 10^4$.
The dynamic range can be improved in several ways.
One method is to employ a smaller cutoff frequency $f_c$, which will impact RBW.
Alternatively, the thickness $d$ of the dielectric layer can be adjusted to affect the desired change in dynamic range.
Care must be taken to ensure that $U_{\rm lpf}$ remains larger than the MDS for the entirety of the scanning bandwidth $\Delta f$.
The dynamic range is independent of the scan rate $\rho$.

The RBW of the simulated spectrum analysis matches well with the theoretical RBW in (\ref{rbw}).
Specifically, the average full width half maximum of the individual spikes in Fig. \ref{spectrum}b was $\approx 200$~MHz, which is near the theoretical limit for RBW from (\ref{rbw}).

For the simulated parameters, the maximum ramp rate is $\rho_{\rm max}\approx 0.1$~THz/ps, several orders of magnitude faster than simulated here.
The RBW is of course dependent on $\rho$ according to (\ref{rbw}), and the frequency sensitivity is relatively unchanged with increasing ramp rate, while for phase dependent variation in the peak amplitude, the relative error increases with increasing RBW.

We also wish to note that this algorithm does not requires that the auto-oscillator frequency, $f_R(t)$, be a perfect linear function of time as in equation (\ref{eq:fr}).
For a system with a scan rate so fast that the ATJ has strongly non-linear frequency behavior, a different matching filter may be required.
Nonetheless, the matching filter $h(f_m,t)$ presented above, in practice, can perform robustly enough to allow for normal levels of noise and intrinsic non-linear frequency dependance on $I_{\rm drive}$.

\section{Conclusion}
\label{s:Conclusion}

In conclusion, we have presented theory that a Pt/AFM/MgO/Pt ATJ can generate an oscillating TAMR with THz frequency.
It was demonstrated with simulation that these THz frequency TAMR oscillations can be dynamically tuned to increase linearly with time at rates as fast as 0.1~THz/ps.
A circuit model was presented that allowed the optimization of the ATJ output voltage by adjusting the thickness and area of the MgO layer, thus allowing impedance matching between an ATJ and an external THz signal to be improved.
We then presented a basic THz signal detector, and a THz spectrum analyzer.
The spectrum analyzer, as simulated with optimized parameters, scanned between 0.5~THz and 0.75~THz in 25 ns, with a dynamic range of $10^4$, a resolution bandwidth near the theoretical limit, and determined the frequency of an external signal with a relative error less than $10^{-4}\%$.

\section*{Acknowledgments}
This work was supported by the Grants Nos. EFMA-1641989 and ECCS-1708982 from the NSF of the USA, and by the DARPA M3IC Grant under the Contract No. W911-17-C-0031.
The publication contains the results of studies conducted by President's of Ukraine grant for competitive projects (F 84).
This work was also supported in part by the grants Nos. 18BF052-01M and 19BF052-01 from the Ministry of Education and Science of Ukraine and the grant No. 1F from the National Academy of Sciences of Ukraine.


\end{document}